 \documentclass[preprint,onecolumn,amsmath,amssymb,aps,prb]{revtex4-1}

\usepackage{graphicx}% Include figure files
\usepackage{dcolumn}% Align table columns on decimal point
\usepackage{bm}% bold math
 \usepackage{float} 
\usepackage{fullpage}
\usepackage{color}
\usepackage{subcaption}
\usepackage{multirow}
\usepackage{amsmath}
\usepackage{amssymb}
\usepackage{amsthm}
\usepackage[colorlinks   = true, urlcolor     = cyan, linkcolor    = blue, citecolor   = red]{hyperref}
\usepackage{bm}
\usepackage{pict2e}
\usepackage{epstopdf}
\usepackage{comment}
\usepackage{epsfig}
\usepackage{url}
\usepackage[ruled,vlined]{algorithm2e}

\begin{document}

%\preprint{APS/123-QED}

\title{Torsional strain engineering of transition metal dichalcogenide nanotubes: An ab initio study}

\author{Arpit Bhardwaj}
\affiliation{College of Engineering, Georgia Institute of Technology, Atlanta, GA 30332, USA}

\author{Abhiraj Sharma}
\affiliation{College of Engineering, Georgia Institute of Technology, Atlanta, GA 30332, USA}

\author{Phanish Suryanarayana}
\email{phanish.suryanarayana@ce.gatech.edu}
%\homepage{https://www.phanishgroup.com}
%\email{phanish.suryanarayana@ce.gatech.edu}
\affiliation{College of Engineering, Georgia Institute of Technology, Atlanta, GA 30332, USA}

%\date{\today}

\begin{abstract}
We study the effect of torsional deformations on the electronic properties of single-walled transition metal dichalcogenide (TMD) nanotubes.  In particular, considering forty-five select armchair and zigzag TMD nanotubes, we perform symmetry-adapted Kohn-Sham density functional theory calculations to determine the variation in bandgap and effective mass of charge carriers with twist. We find that metallic nanotubes remain so even after deformation, whereas semiconducting nanotubes experience a decrease in bandgap with twist --- originally direct bandgaps become indirect --- resulting in semiconductor to metal transitions.  In addition, the effective mass of  holes and electrons continuously decrease and increase with twist, respectively,  resulting in  n-type to p-type semiconductor transitions. We find that this behavior is likely due to rehybridization of orbitals in the metal and chalcogen atoms, rather than charge transfer between them. Overall, torsional deformations represent a powerful avenue to engineer the electronic properties of semiconducting TMD nanotubes, with applications to devices like sensors and semiconductor switches. 
\end{abstract}

%\keywords{Torsional modulus, Transition Metal Dichalcogenides, Nanotubes, Density Functional Theory, Shear modulus, Young's modulus, Poisson's ratio}

\maketitle

%%%MAIN TEXT%%%%
%%%%%%%%%%%%%%%%%%%%

%%%%%%%%%%%%%%%%%%%%%%%%%%%%%%%
%%%   Introduction %%%%%%%%%%%%%%%%%
%%%%%%%%%%%%%%%%%%%%%%%%%%%%%%%
The synthesis of carbon nanotubes three decades ago \cite{iijima1991helical} represents a pioneering contribution that has  played a pivotal role in the creation of the  now ever-expanding  field of nanoscience/nanotechnology. The significant progress in this field is epitomized by the nearly  two dozen different nanotubes that have been synthesized to date \cite{tenne2003advances, rao2003inorganic, serra2019overview}, with the potential for many more in the future, given that thousands of  their two-dimensional analogues have been predicted to be stable from ab initio  calculations \cite{haastrup2018computational, zhou20192dmatpedia, gjerding2021recent}. Nanotubes have been extensively researched, inspired by the novel and magnified electronic, mechanical, thermal, and optical properties, relative to their bulk counterparts \cite{tenne2003advances, rao2003inorganic, serra2019overview}. In particular, a number of strategies have been developed to tune/engineer these properties, including chirality/radius \cite{ghosh2019symmetry, maiti2003bandgap, zhang2011helical, guo2005systematic, jia2006structure, yang2005electronic, mintmire1993properties, wang2017band, ding2002analytical, ouyang2002fundamental}, defects\cite{wang2019investigation, parashar2015switching, akdim2011bandgap}, electric field\cite{tien2005band, chen2004band, akdim2011bandgap}, and mechanical deformations\cite{yang2000electronic, rochefort1999electrical, tombler2000reversible, yang1999band, maiti2003bandgap, kim2001electronic, kinoshita2010electronic, coutinho2009band, ghassemi2012field}, highlighting the technological importance of nanotubes. 

Among the various categories of nanotubes --- classified based on the two-dimensional material from which they can be thought to arise based on a roll-up construction --- the transition metal dichalcogenide (TMD) group, which has materials of the form MX\textsubscript{2}, where M and X represent a transition metal and chalcogen, respectively,  is  the most diverse, with the largest number of distinct  nanotubes  synthesized to date \cite{tenne2003advances, rao2003inorganic, serra2019overview}. This manifests itself into varying electronic properties encompassing semiconducting \cite{seifert2000structure, seifert2000electronic}, metallic \cite{seifert2000novel, enyashin2005computational}, and superconducting \cite{nath2003superconducting, tsuneta2003formation}. Notably, a number of mechanisms have been found to tune/tailor these properties, including chirality/radius \cite{ivanovskaya2003computational, ivanovskaya2004tubular, gao2017structural, yin2016chiral, milovsevic2007electronic, teich2011structural, scheffer2002scanning, zibouche2012layers, seifert2000electronic, ansari2015ab, seifert2000structure}, defects \cite{tal2001effect, li2015tailoring}, temperature \cite{nath2003superconducting, tsuneta2003formation}, electric field \cite{wang2014tuning, zibouche2019strong}, and mechanical deformation \cite{zibouche2014electromechanical, oshima2020geometrical, li2014strain, wang2016strain, ghorbani2013electromechanics, lu2012strain, levi2015nanotube}. This makes TMD nanotubes ideally suited for a number of technological applications, including nanoelectromechanical (NEMS) devices \cite{yudilevichself, levi2015nanotube, divon2017torsional}, photodetectors \cite{unalan2008zno, zhang2012high, zhang2019enhanced}, mechanical sensors\cite{li2016low, sorkin2014nanoscale, oshima2020geometrical}, biosensors\cite{barua2017nanostructured}, and superconductive materials \cite{nath2003superconducting, tsuneta2003formation}.

Strain engineering represents an elegant and efficient way to control the electronic properties of TMD nanotubes, as shown experimentally \cite{levi2015nanotube} as well as theoretically from ab initio Kohn-Sham density functional theory (DFT) calculations \cite{zibouche2014electromechanical, oshima2020geometrical, li2014strain, wang2016strain, ghorbani2013electromechanics, lu2012strain}. However, other than the experimental work referenced above,  where the effect of both tensile and torsional deformations have been studied for the WS\textsubscript{2} nanotube, research efforts have focused solely on tensile/compressive deformations, and that too for only a small fraction of the materials in the TMD nanotube group. Indeed, the study of torsional deformations at practically relevant twists and nanotube diameters requires large number of atoms when employing the standard periodic boundary conditions \cite{sharma2021real}, which makes it intractable to first principles methods like Kohn-Sham DFT, given its cubic scaling with system size and large associated prefactor. Therefore, the electromechanical response of TMD nanotubes to torsional deformations has remained unexplored heretofore, providing the motivation for the current work. 

In this work, utilizing a  recently developed symmetry-adapted formulation and implementation for Kohn-Sham DFT \cite{sharma2021real}, we perform a comprehensive study of the electronic response of single-walled TMD nanotubes to torsional deformations.  In particular, we determine the variation in bandgap and charge carriers' effective mass at practically relevant twists and diameters for forty-five select armchair and zigzag TMD nanotubes. We also provide fundamental insights into the observed behavior. 

%%%%%%%%%%%%%%%%%%%%%%%%%%%%%%%
%%%%%%%%%%%%%%%%%%%%%%%%%%%%%%%

%%%%%%%%%%%%%%%%%%%%%%%%%%%%%%%
%%%  Systems and methods  %%%%%%%%%%
%%%%%%%%%%%%%%%%%%%%%%%%%%%%%%%

We consider the following TMD nanotubes with 1T-o symmetry \cite{nath2002nanotubes, bandura2014tis2}: M$=$\{Ti, Zr, Hf, Mn, Ni, Pd, Pt\} and X$=$\{S, Se, Te\}; and the following ones with 2H-t symmetry \cite{nath2002nanotubes, bandura2014tis2}: M$=$\{V, Nb, Ta, Cr, Mo, W, Fe, Cu\} and X$=$\{S, Se, Te\}. The current list includes all materials that have been synthesized as single/multi-walled nanotubes \cite{nath2001mose2, nath2001simple, chen2003titanium, nath2001new,nath2002nanotubes, tenne1992polyhedral, gordon2008singular, bruser2014single, remskar2001self}, and/or those that have stable two-dimensional monolayer counterparts, as predicted from ab initio calculations \cite{haastrup2018computational, heine2015transition, guo2014tuning}. We choose the nanotube radii --- values provided in Supplementary Information --- to be commensurate with those that are synthesized, and in cases where this is yet to happen, the radii are chosen to be commensurate with synthesized nanotubes that are expected to have similar structure.

\begin{figure}[h!]
        \centering
        \includegraphics[width=0.8\textwidth]{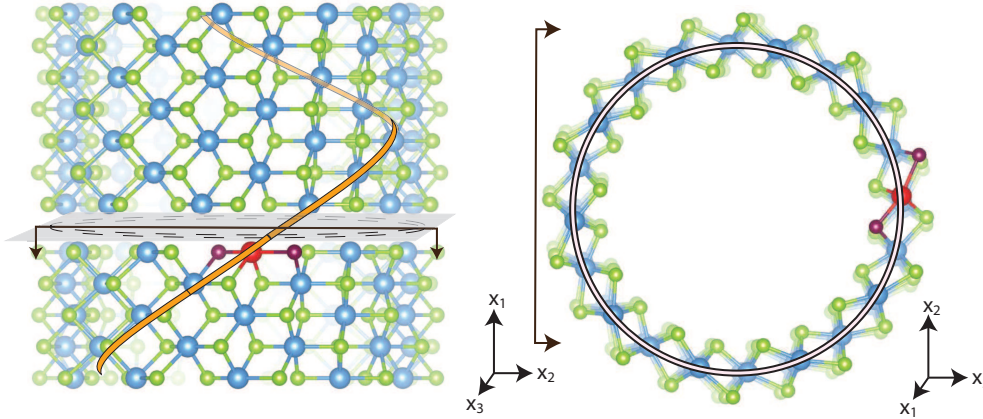}
        \caption{Illustration depicting the cyclic and helical symmetry inherent to a twisted (10,10) 1T-o TMD nanotube (structural model generated using VESTA \cite{momma2008vesta}).  In particular, all atoms can be considered to be cyclic and/or helical images of the metal and chalcogen atoms colored red and maroon, respectively. This symmetry is exploited while performing ab initio Kohn-Sham calculations using the Cyclix-DFT code \cite{sharma2021real, xu2021sparc}.}
      \label{fig:illustration}
    \end{figure}

TMD nanotubes have cyclic and helical symmetry inherent to their structure, which remains the case even after the application of torsional deformations, as illustrated in Fig.~\ref{fig:illustration}. We exploit this feature using the recently developed Cyclix-DFT code \cite{sharma2021real}, which represents an implementation of the cyclic+helical symmetry-adapted formulation \cite{sharma2021real, ghosh2019symmetry, banerjee2016cyclic} for the Kohn-Sham problem within the large-scale parallel real-space DFT code SPARC \cite{xu2021sparc, ghosh2017sparc1, ghosh2017sparc2}. This allows the use of a fundamental domain containing only one metal and two chalcogen atoms,  as illustrated in Fig.~\ref{fig:illustration}, thereby tremendously reducing the cost of the calculations. Indeed, such simulations are beyond the reach of even state-of-the-art DFT codes \cite{banerjee2018two, motamarri2020dft, xu2021sparc} when employing the traditional periodic boundary conditions, e.g., a (60,60) TiS\textsubscript{2} nanotube with diameter $\sim 10$ nm and an external twist of $15\times10\textsuperscript{-4}$ rad/Bohr has $194,760$ atoms in the simulation domain, a system size that is intractable even on large-scale supercomputers. It is worth noting that the Cyclix-DFT code has already proven to be an effective and  reliable tool in a number of  physical applications \cite{PhysRevMaterials.5.L030801, kumar2021flexoelectricity, sharma2021real, kumar2020bending, bhardwaj2021torsional}, providing further evidence of the fidelity of the computations performed here. 

In all simulations, we employ the semilocal Perdew–Burke–Ernzerhof (PBE) \cite{perdew1996generalized}  exchange-correlation functional and optimized norm-conserving Vanderbilt (ONCV)  pseudopotentials \cite{hamann2013optimized} from the SG15 \cite{SCHLIPF201536} collection. In addition to the developer tests \cite{SCHLIPF201536}, the transferability of the   pseudopotentials has been verified by comparisons with  all-electron DFT code Elk \cite{elk} for select bulk systems, and by the  results obtained in recent work \cite{bhardwaj2021torsional, kumar2021flexoelectricity, kumar2020bending}. In particular, the equilibrium configuration for the nanotubes and their two-dimensional counterparts \cite{bhardwaj2021torsional} are in very good agreement with previous DFT results employing the same exchange-correlation functional \cite{haastrup2018computational, zhou20192dmatpedia, wang2016strain, xiao2014theoretical, li2014strain,ataca2012stable, chang2013orbital, amin2014strain, guo2014tuning}. Furthermore, the geometries are also in very good agreement with experimental measurements for both the nanotubes \cite{chen2003titanium, nath2003superconducting, nath2002nanotubes, nath2001mose2} and their two-dimensional analogues \cite{klots2014probing, ugeda2014giant, hill2016band, novoselov2005two, coleman2011two}, justifying the choice of PBE exchange-correlation functional in this work. Indeed, PBE is known to under-predict the bandgap of TMD monolayers --- expected to have similar band structure as the nanotubes, given that they have considerably large diameters where curvature effects are  minor --- relative to hybrid functionals like HSE \cite{haastrup2018computational}. However, there is good agreement in the overall band structure  and nature of bandgap \cite{haastrup2018computational}. In particular,  we are interested in general trends, which are expected to be insensitive to the choice of exchange-correlation functional, particularly given the small twists considered here. Even quantitatively, hybrid functionals are not necessarily more accurate than PBE in predicting the band structure, e.g., bulk TMDs \cite{kuc2011influence}. In view of this and  the tremendously larger cost associated with hybrids, PBE has been the functional of choice for TMD nanotubes \cite{ghorbani2013electromechanics, zibouche2012layers, zibouche2019strong, wang2016strain,  zibouche2014electromechanical, yin2016chiral, wang2014tuning, li2015tailoring, li2014strain, lu2012strain, ansari2015ab}. Note that the incorporation of spin-orbit coupling (SOC) causes relatively minor modifications to the band structure \cite{haastrup2018computational}, which is why it has been neglected here.

We perform the  symmetry-adapted Kohn-Sham DFT calculations described above to determine the variation in bandgap and effective mass of charge carriers (i.e., electrons and holes) with shear strain for the forty-five select armchair and zigzag TMD nanotubes. The shear strain is defined to be the product of the nanotube radius and the applied twist per unit length. The values for shear strain are chosen to be commensurate with those found in experiments \cite{levi2015nanotube, divon2017torsional, nagapriya2008torsional}. Additional details regarding the calculation of the bandgap and effective mass within the symmetry-adapted formulation can be found in previous work \cite{sharma2021real}. The numerical parameters in Cyclix-DFT, including real-space grid spacing,  Brillouin zone integration grid spacing, vacuum in the radial direction, and structural relaxation tolerances (both cell and atom) are chosen such that the bandgap and effective mass are calculated to within an accuracy of 0.01 eV and 0.01 a.u., respectively. This translates to the requirement of the ground state energy being converged to within $10^{-4}$ Ha/atom, respectively. The simulation data for all the results presented below can be found in the Supplementary Information. 

In Fig.~\ref{fig:heatbandgap}, we present the variation of the bandgap with shear strain for the selected TMD nanotubes. We observe that the untwisted MoS\textsubscript{2}, MoSe\textsubscript{2}, MoTe\textsubscript{2}, WS\textsubscript{2}, WSe\textsubscript{2}, WTe\textsubscript{2}, CrS\textsubscript{2}, CrSe\textsubscript{2}, CrTe\textsubscript{2}, PdS\textsubscript{2}, PdSe\textsubscript{2}, PdTe\textsubscript{2}, PtS\textsubscript{2}, PtSe\textsubscript{2}, PtTe\textsubscript{2}, ZrS\textsubscript{2}, ZrSe\textsubscript{2}, HfS\textsubscript{2}, HfSe\textsubscript{2}, NiS\textsubscript{2}, NiSe\textsubscript{2}, and TiS\textsubscript{2} nanotubes are semiconducting, while the remaining are metallic. In addition, nanotubes that are metallic continue to be so even after the application of twist, whereas semiconducting nanotubes undergo a decrease in bandgap value with twist --- bandgaps that are originally direct become indirect ---  resulting in a semiconductor to metal transition. In particular, armchair HfSe\textsubscript{2}, ZrSe\textsubscript{2}, PtTe\textsubscript{2}, NiS\textsubscript{2}, TiS\textsubscript{2}, NiSe\textsubscript{2}, and PdTe\textsubscript{2}; and zigzag ZrSe\textsubscript{2}, PtTe\textsubscript{2}, NiS\textsubscript{2}, TiS\textsubscript{2}, NiSe\textsubscript{2}, and PdTe\textsubscript{2} nanotubes undergo a semiconductor to metal transition for the twists considered. The transition for TiS\textsubscript{2}, NiSe\textsubscript{2}, and PdTe\textsubscript{2} nanotubes occurs at substantially  lower strains than the others, since the bandgaps in the untwisted state are smaller.  Such transitions are also expected for the remaining semiconducting nanotubes, however the amount of twist required to achieve this can be significantly higher, at which point stability considerations become particularly important. Tunability of the bandgap and controlled semiconductor-metal transitions like those observed here have applications in devices such as  mechanical sensors\cite{li2016low, sorkin2014nanoscale, oshima2020geometrical}.

\begin{figure}[h!]
        \centering
        \includegraphics[width=0.95\textwidth]{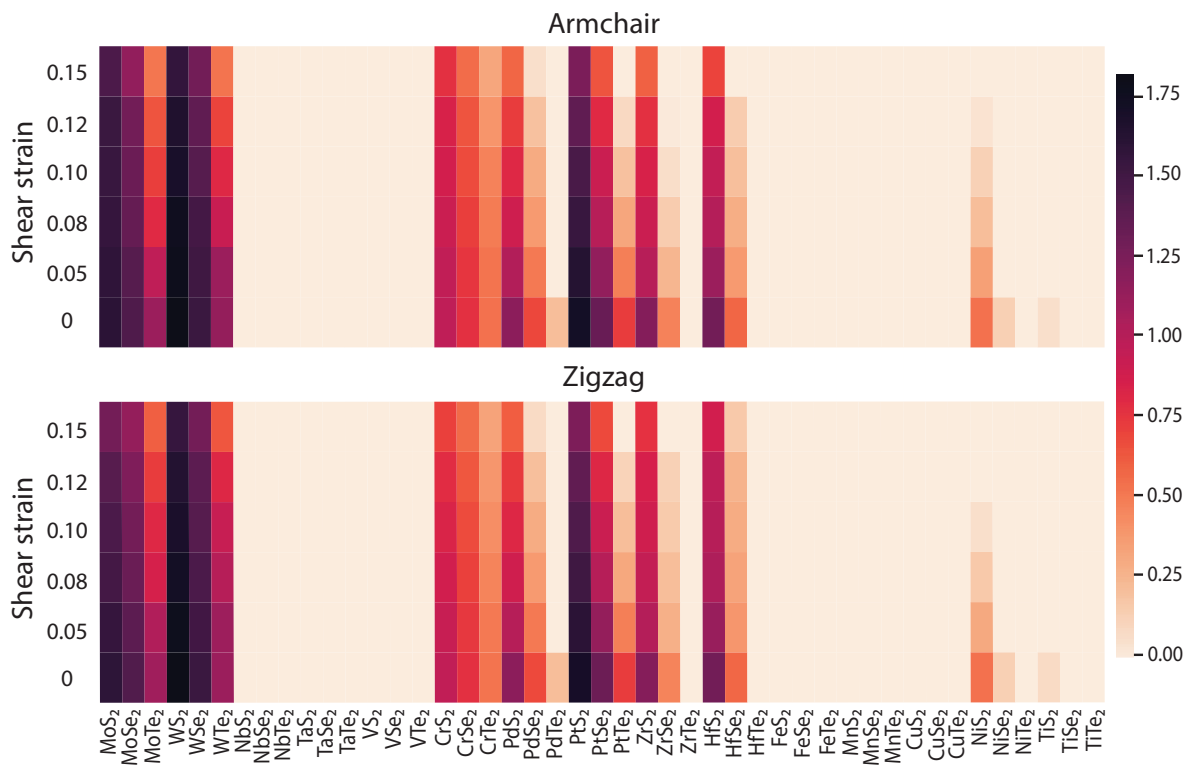}
        \caption{Variation of bandgap with twist for the forty-five select armchair and zigzag TMD nanotubes.}
      \label{fig:heatbandgap}
    \end{figure}

\begin{figure}[htbp!]
        \centering
        \includegraphics[width=0.95\textwidth]{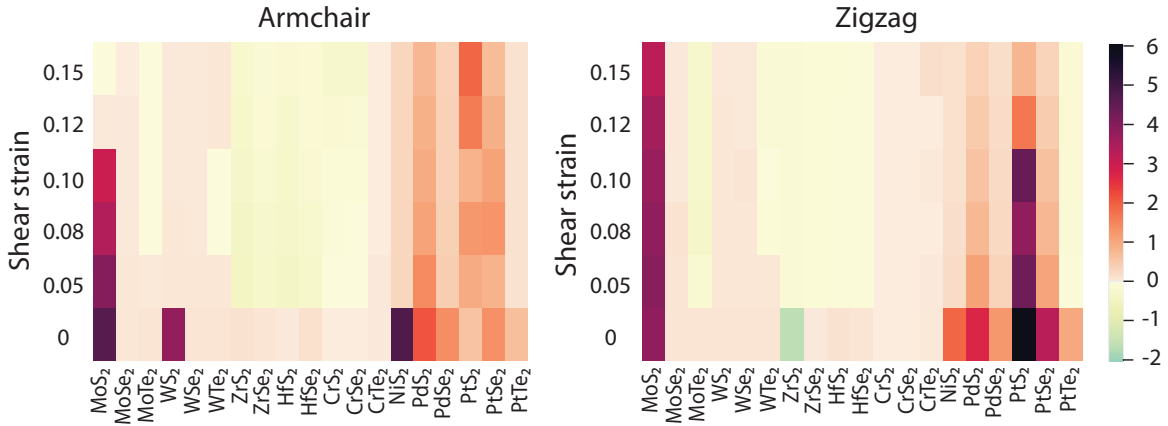}
        \caption{Variation of the difference in effective mass between holes  and electrons (holes minus electrons) with twist for the nineteen semiconducting armchair and zigzag TMD nanotubes. }
      \label{fig:heatholemass}
    \end{figure}

In Fig.~\ref{fig:heatholemass}, we present the variation of the difference in effective mass between holes  and electrons  with twist for the nineteen semiconducting armchair and zigzag TMD nanotubes that were identified above. The effective mass of the holes and electrons relative to each other  can be used to identify whether the nanotubes are n-type or p-type semiconductors. Specifically, the effective mass of the holes being greater than electrons suggests that the electrons have higher mobility, resulting in n-type semiconductors, with the reverse being true for p-type semiconductors.  It is clear from the figure that other than the zigzag ZrS\textsubscript{2} nanotube, all other nanotubes are n-type semiconductors in their untwisted state. Upon the application of twist, the effective mass of the holes continuously decreases while that of the electrons continuously increases, leading to a crossover in their values. In particular, armchair MoS\textsubscript{2}, MoTe\textsubscript{2}, WTe\textsubscript{2}, ZrS\textsubscript{2}, ZrSe\textsubscript{2}, HfS\textsubscript{2}, HfSe\textsubscript{2}, CrS\textsubscript{2}, and CrSe\textsubscript{2}; and zigzag MoTe\textsubscript{2}, WTe\textsubscript{2}, ZrSe\textsubscript{2}, HfS\textsubscript{2}, HfSe\textsubscript{2}, CrS\textsubscript{2}, CrSe\textsubscript{2}, and PtTe\textsubscript{2} nanotubes undergo a transition from n-type to p-type semiconducting behavior for the twists considered. Indeed, larger twists are likely to result in transitions for the other nanotubes as well, however, as mentioned above, stability considerations become particularly important in such scenarios. Controlled n-type to p-type semiconductor transitions like those observed here have applications in semiconductor switches\cite{nilges2009reversible, zhang2016switching, hiramatsu2007heavy, chen2015p, wen2018pressure, naab2013high}. 

The results presented in this work are in good agreement with those available in literature. Specifically, in the untwisted state, the metallic nature predicted for TaS\textsubscript{2}, NbSe\textsubscript{2},  and NbS\textsubscript{2} nanotubes is in agreement with tight binding calculations \cite{enyashin2005computational, enyashin2004interatomic,  ivanovskaya2003electronic, seifert2000novel}; the bandgap values for MoS\textsubscript{2}, MoSe\textsubscript{2}, WS\textsubscript{2}, WSe\textsubscript{2}, and CrS\textsubscript{2} nanotubes are in good agreement with other DFT studies  \cite{wang2016strain, zibouche2014electromechanical}; and the effective masses of electrons for MoS\textsubscript{2}, WS\textsubscript{2}, and CrS\textsubscript{2} nanotubes are in good agreement with other DFT results \cite{wang2016strain, li2014strain, zibouche2014electromechanical}. In addition, the bandgap variation upon twisting for WS\textsubscript{2}  and MoS\textsubscript{2} nanotubes is in good qualitative agreement with previous experiments  and tight binding calculations, respectively \cite{levi2015nanotube}.  A quantitative comparison cannot be made due to availability of only  electrical response in the experiments, and the diameters in both experiments and tight binding simulations being different to those chosen here. Indeed, we have found that the bandgap variation with shear strain is qualitatively similar for different diameters. In view of this, we note that the ratio of change in bandgap between 15$\%$ and 10$\%$ shear strains for WS\textsubscript{2} nanotube is 2.1 based on experiments \cite{levi2015nanotube}, which is in good agreement with the ratio of 1.9 obtained here for both armchair and zigzag variants. Furthermore, the ratio of bandgap between 15\% and 0\% strains for the armchair MoS\textsubscript{2} nanotube is 0.9 based on tight binding results \cite{levi2015nanotube}, which is in excellent agreement with the value of 0.9 obtained here. It is interesting to note that metallic TMD nanotubes continue to be so even after the application of torsional deformations, which is fundamentally different from the response of carbon nanotubes \cite{sharma2021real}.

\begin{figure}[htbp!]
        \centering
        \includegraphics[width=0.88\textwidth]{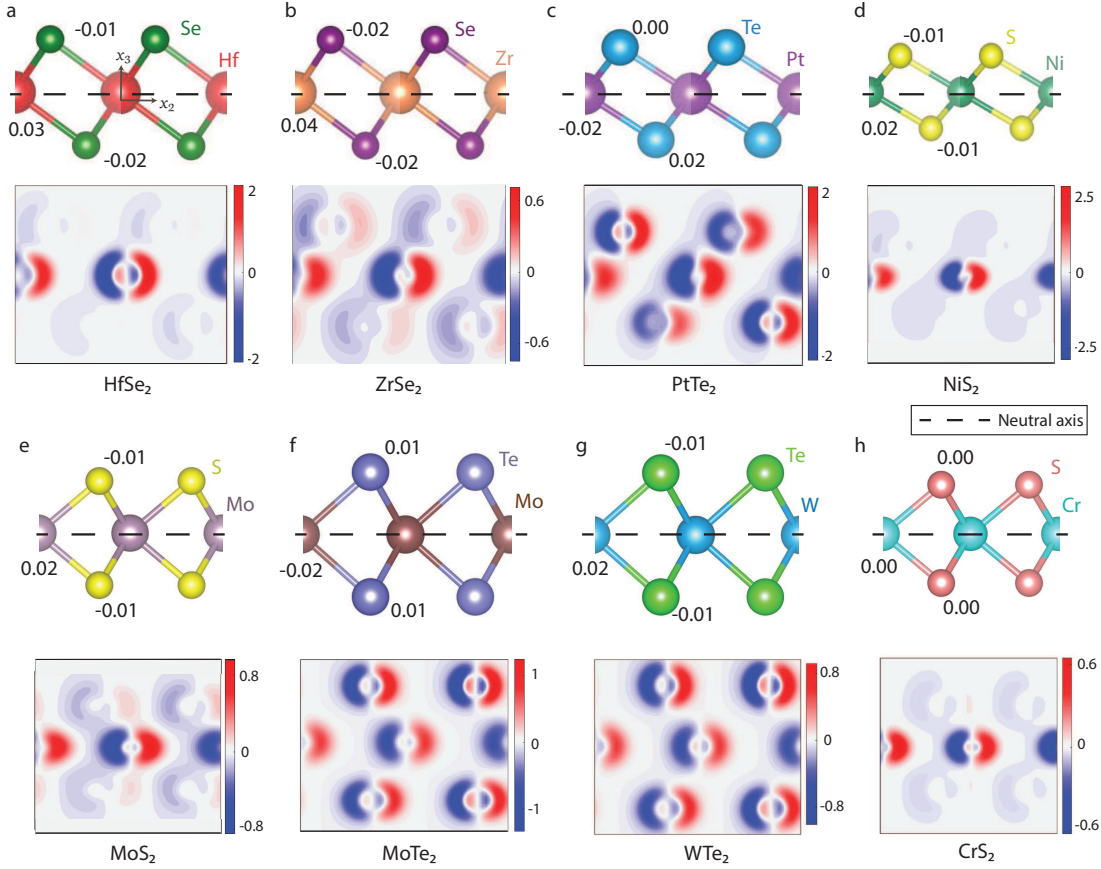}
        \caption{Contours of electron density difference --- integrated along the $x_1$ direction --- between the twisted and untwisted armchair nanotube configurations (twisted minus untwisted). The twists chosen for a, b, c, and d correspond to the semiconductor--metal transition; and in e, f, g, and h they correspond to the transition from n-type to p-type semiconductors. The contours are plotted on the $x_2 x_3$-plane in the corresponding monolayer flat sheet configuration. The charge transfer due to twisting, shown near the corresponding atoms in the lattice structure, is obtained from Bader analysis \cite{bader1981quantum, tang2009grid}.}
      \label{fig:contour}
    \end{figure}

To gain further insights into the results presented above, choosing representative TMD nanotubes that  demonstrate  semiconductor to metal and n-type to p-type transitions, we plot the contours of electron density difference  between the twisted and untwisted nanotube configurations in Fig.~\ref{fig:contour}.  In addition, we compute the charge transfer due to torsional deformations using Bader analysis \cite{bader1981quantum, tang2009grid}. We observe that there is negligible change in the Bader charge, suggesting the lack of charge transfer between the metal and chalcogen atoms, indicating that the nature of bonding between them remains unchanged. It can therefore be inferred that electronic variations due to torsional deformations, as visible through the change in electron density contours, is likely due to the rehybridization of orbitals in the metal and chalcogen atoms. 

%%%%%%%%%%%%%%%%%%%%%%%%%%%%%%%
%%%  Conclusions  %%%%%%%%%%%%%%%%%
%%%%%%%%%%%%%%%%%%%%%%%%%%%%%%%

In summary, we have studied the electronic response of single-walled TMD nanotubes to torsional deformations. In particular, using symmetry-adapted first principles DFT simulations, we have  determined the variation in bandgap and effective mass of charge carriers with twist for forty-five select armchair and zigzag TMD nanotubes. We have found that whereas the nature of originally metallic nanotubes remains unaltered, there is a continuous decrease in bandgap --- changes to indirect for systems that are originally direct --- with increasing twist for semiconducting TMD nanotubes, culminating in semiconductor to metal transitions. In addition, we have found that the effective mass of  holes and electrons continuously decrease and increase with twist, respectively,  culminating in  transitions from n-type to p-type semiconducting  behavior.  We have found that these changes can be attributed to rehybridization of orbitals in the metal and chalcogen atoms, rather than charge transfer between them. Overall, we conclude that torsional deformations represent a powerful tool to tailor the electronic properties of semiconducting TMD nanotubes, with applications to devices such as sensors and semiconductor switches.

The current work suggests a number of interesting directions for future research. These include studying the electromechanical response of multi-walled TMD nanotubes, which  are  of significant practical relevance given their ease of synthesis; and extension of such studies to large twists, where unexpected nonlinear behavior can be observed.

%%%%%%%%%%%%%%%%%%%%%%%%%%%%%%%
%%%%%%%%%%%%%%%%%%%%%%%%%%%%%%%
%%%%%%%%%%%%%%%%%%%%%%%%%%%%%%%%%

\section*{Acknowledgments}
The authors gratefully acknowledge the support of the US National Science Foundation (CAREER–1553212).
 \vspace{-1mm}

\bibliographystyle{unsrt}
%\bibliography{refer.bib}

\end{document}